\documentclass[twocolumn,prd,groupaddress,
showpacs,floatfix,preprintnumbers,nofootinbib]{revtex4}

\usepackage{mathrsfs}
\usepackage{epsfig}
\usepackage{amsmath,bm}
\usepackage{slashed}

\begin{document}

\preprint{MPP-2011-21}

\title{Supernova bound on keV-mass sterile neutrinos reexamined}%

\author{Georg G. Raffelt}%
%\email{raffelt@mppmu.mpg.de}

\author{Shun Zhou}%
%\email{zhoush@mppmu.mpg.de}

\affiliation{Max-Planck-Institut f\"{u}r Physik
  (Werner-Heisenberg-Institut), F\"{o}hringer Ring 6, D-80805
  M\"{u}nchen, Germany}

\date{25 February 2011}%

%%%%%%%%%%%%%%%%%%%%%%%%%%%%%%%%%%%%%%%%%%%%%%%%%%%%%%%%%%%%%%%%%%%%%%
\begin{abstract}
Active-sterile neutrino mixing is strongly constrained for $m_s \agt
100$~keV to avoid excessive energy losses from supernova cores. For
smaller $m_s$, matter effects suppress the effective mixing angle
except for a resonant range of energies where it is enhanced. We
study the case of $\nu_\tau$-$\nu_s$ mixing where a
$\nu_\tau$-$\bar\nu_\tau$ asymmetry builds up due to the strong
excess of $\nu_s$ over $\bar\nu_s$ emission or vice versa, reducing
the overall emission rate. In the warm dark matter range $m_s \alt
10$~keV the mixing angle is essentially unconstrained.
\end{abstract}
%%%%%%%%%%%%%%%%%%%%%%%%%%%%%%%%%%%%%%%%%%%%%%%%%%%%%%%%%%%%%%%%%%%%%%


\maketitle

%%%%%%%%%%%%%%%%%%%%%%%%%%%%%%%%%%%%%%%%%%%%%%%%%%%%%%%%%%%%%%%%%%%%%%
\section{Introduction}                        \label{sec:introduction}
%%%%%%%%%%%%%%%%%%%%%%%%%%%%%%%%%%%%%%%%%%%%%%%%%%%%%%%%%%%%%%%%%%%%%%

Sterile neutrinos $\nu_s$ can be produced in the early universe or
in supernova (SN) cores if they mix with one of the active
flavors~\cite{Kusenko:2009up,Boyarsky:2009ix}. Even if the mixing
angle $\theta$ is very small, repeated collisions of the active
component allow for an efficient $\nu_s$ production. The SN~1987A
neutrino signal duration implies $\sin^22\theta \alt 10^{-9}$ to
avoid excessive energy
losses~\cite{Kainulainen:1990bn,Raffelt:1992bs,Peltoniemi:1992,
Shi:1993ee,Dolgov:2000ew,Dolgov:2000jw,Dolgov:2000pj,Abazajian:2001nj}.
In detail, this limit depends on whether the dominant mixing is with
$\nu_e$ or one of the other active flavors. This result assumes $m_s
\agt 100$~keV for matter effects on neutrino propagation to be
negligible compared with the vacuum mass.

For smaller masses, the matter effect typically suppresses the
effective mixing angle and thus diminishes the limit on
$\sin^22\theta$. For $m_s \alt 1$~keV even maximal mixing is
allowed. However, no detailed treatment of the SN bound exists for
$1~{\rm keV} \alt m_s \alt 100~{\rm keV}$, where matter and
resonance effects are important~\cite{Raffelt:1992bs,Shi:1993ee}. On
the other hand, this is precisely the mass range where sterile
neutrinos could play an interesting warm dark matter role in
cosmology~\cite{Dodelson:1993je,Shi:1998km,Asaka:2006nq}. While
sterile neutrinos can be produced in the early universe by different
mechanisms \cite{Dodelson:1993je,Shi:1998km}, the production by
oscillations and collisions once more depends on the active-sterile
mixing angle, so one naturally wonders about the SN bound on
$\theta$.

In particular, we study the role of feedback of sterile neutrino
emission on the emission rate itself. When matter effects are
important, the mixing angle is resonantly enhanced in some range of
neutrino
energies~\cite{Wolfenstein:1977ue,Mikheyev:1986gs,Kuo:1989qe}. At
first the emission of $\bar\nu_s$ is more efficient than $\nu_s$
because in the neutrino sector the mixing angle is suppressed for
all energies. As a consequence, the active flavor, being trapped in
the SN core, builds up a $\nu$ excess. The opposite could happen for
larger mixing angles, when most antineutrinos $\bar\nu_s$ are
trapped rather than freely escape and the emission of $\nu_s$ is
more efficient than that of $\bar\nu_s$. To be specific we use
$\nu_\tau$ as the active flavor because there is no initial
$\nu_\tau$-$\bar\nu_\tau$ asymmetry in a SN core and $m_\tau$ is so
large that charged $\tau$ leptons never play any role. Our main
point is that the depletion of $\nu_\tau$ or $\bar\nu_\tau$ relative
to the other always goes in the direction of quenching the initial
emission rate, implying that in the 1--100~keV-mass range the
$\theta$ bounds are indeed suppressed.

Our conclusion differs somewhat from previous discussions where it
was suggested that the buildup of a $\nu_\tau$-$\bar\nu_\tau$
asymmetry goes in the direction of reducing the matter effect and
leads to restrictive SN limits on $\theta$~\cite{Abazajian:2001nj}.
While we agree that the matter effect can be modified in this
direction, the positive $\nu_\tau$-$\bar\nu_\tau$ asymmetry also
implies a depletion of source $\bar\nu_\tau$ to be converted to
$\bar\nu_s$ relative to source $\nu_\tau$ to be converted to
$\nu_s$. It is also possible in some region of the parameter space
that the matter effects are enlarged due to a negative
$\nu_\tau$-$\bar\nu_\tau$ asymmetry. In both cases, the compound
effect is a reduction, not an enhancement, of the energy loss.

In Sec.~\ref{sec:matter}, we briefly review the matter effects on
active-sterile neutrino mixing in the SN core. The development of a
$\nu_\tau$-$\bar\nu_\tau$ asymmetry is discussed in
Sec.~\ref{sec:state}, where we identify a stationary state with equal
neutrino and antineutrino emission rates and estimate the time scale
to reach it. Sec.~\ref{sec:snbounds} is devoted to the calculation of
the energy-loss rate caused by sterile neutrinos and the SN bound on
sterile neutrino masses and mixing angles. Finally, we summarize our
conclusions in Sec.~\ref{sec:conclusions}.

%%%%%%%%%%%%%%%%%%%%%%%%%%%%%%%%%%%%%%%%%%%%%%%%%%%%%%%%%%%%%%%%%%%%%
\section{Matter effects}                         \label{sec:matter}
%%%%%%%%%%%%%%%%%%%%%%%%%%%%%%%%%%%%%%%%%%%%%%%%%%%%%%%%%%%%%%%%%%%%%

The dispersion relation of neutrinos will be modified in matter due
to the coherent forward scattering of neutrinos off background
particles~\cite{Kuo:1989qe}. This matter effect can be described by
an effective potential $V_{\nu_\alpha}$ for each kind of active
neutrino $\nu_\alpha = \nu_e$, $\nu_\mu$ and $\nu_\tau$. The
effective potentials for antineutrinos have the opposite signs,
i.e., $V_{\bar\nu_\alpha} = - V_{\nu_\alpha}$. In the case of
$\nu_\tau$-$\nu_s$ oscillation in matter, the effective Hamiltonian
is
\begin{equation}\label{eq:Hamiltonian}
H_{\rm eff} = \left(\begin{matrix} V_{\nu_\tau} - \omega c_{2\theta}
  & \omega s_{2\theta} \cr \omega s_{2\theta} & \omega c_{2\theta} -
  V_{\nu_\tau}\end{matrix} \right)   \; ,
\end{equation}
where $s_{2\theta} \equiv \sin 2\theta$, $c_{2\theta} \equiv \cos
2\theta$ with $\theta$ being the vacuum mixing angle and $\omega
\equiv \Delta m^2/2E$ the oscillation frequency in vacuum. As far as
the keV-mass sterile neutrinos are concerned, we have $\Delta m^2
\approx m^2_s$ with $m_s$ being the sterile neutrino mass. In
contrast with the flavor conversions of ordinary neutrinos in the
SN, there are no collective effects in active-sterile neutrino
oscillations.

In SN cores, the main ingredients of matter are protons $p$,
neutrons $n$, electrons $e$, some muons $\mu$ as well as active
neutrinos $\nu_\alpha$ and antineutrinos $\bar\nu_\alpha$. We ignore
the possibility of a meson condensate or hyperons. In such a medium,
the effective potential for tau neutrinos is
\begin{equation}\label{eq:potential}
V_{\nu_\tau} = \sqrt{2} G_{\rm F} N_{\rm B} \left[-\frac{1}{2} Y_n +
  Y_{\nu_e} + Y_{\nu_\mu} + 2Y_{\nu_\tau}\right]
\; ,
\end{equation}
where $G_{\rm F}$ is the Fermi constant, $N_{\rm B}$ the baryon
number density, and $Y_X \equiv (N_X - N_{\bar X})/N_{\rm B}$ with
$N_X$ and $N_{\bar{X}}$ being the number densities of particle $X$
and its antiparticle $\bar X$. Because of charge neutrality $Y_p =
Y_e + Y_\mu$, we have $Y_n = 1 - Y_e - Y_\mu$. While $\nu_\alpha$
and $\bar\nu_\alpha$ of all flavors can be produced in pairs,
electron neutrinos $\nu_e$ can also be generated in beta processes
such as $e^- + p \to \nu_e + n$, and similar for muon neutrinos. The
muon mass $m_\mu = 106~{\rm MeV}$ is comparable to the average
thermal energy $\langle E \rangle = 3 T$ with $T = 30~{\rm MeV}$,
thus a small population of muons is unavoidable. Beta equilibrium
$\nu_\mu + n \leftrightarrow \mu^- + p$ implies the relation among
chemical potentials $\mu_\mu - \mu_{\nu_\mu} = \mu_n - \mu_p \equiv
\hat{\mu}$, where $\hat{\mu}$ typically lies in the range
50--100~MeV, depending sensitively on the equation of state. Noting
that the initial $\mu$ lepton number is vanishing and taking
$\hat{\mu} = 50~{\rm MeV}$, one finds $\mu_\mu \approx 18~{\rm MeV}$
and $\mu_{\nu_\mu} \approx -32~{\rm MeV}$~\cite{Hannestad:1999zy}.
On the other hand, the $\nu_e$ chemical potential will be much
larger due to the electron lepton number trapped during infall.

For simplicity, we consider a SN core just after the bounce and
assume the temperature $T = 30~{\rm MeV}$ and matter density $\rho =
3.0\times 10^{14}~{\rm g}~{\rm cm}^{-3}$ to be constant. We
furthermore take a typical value of electron lepton number fraction
$Y_L = Y_e + Y_{\nu_e} = 0.37$, which leads to $Y_e = 0.3$ and
$Y_{\nu_e} = 0.07$ due to beta equilibrium with $\hat{\mu} = 50~{\rm
MeV}$. It is straightforward to verify that $Y_\mu/Y_e = 0.01$ and
$Y_{\nu_\mu}/Y_{\nu_e} = -0.05$, so henceforth we simply set $Y_\mu
= Y_{\nu_\mu} = 0$. Tau neutrinos initially follow the Fermi-Dirac
distribution without chemical potential, but later an asymmetry
develops due to $\nu_s$ emission. With these simplifications, the
effective potential in Eq.~(\ref{eq:potential}) is
\begin{equation}\label{eq:potential-simplified}
V_{\nu_\tau} = - \frac{G_{\rm F}}{\sqrt{2}} N_{\rm B} \left(1 -
Y_e - 2Y_{\nu_e} - 4Y_{\nu_\tau}\right) \; .
\end{equation}
It is now evident that $V_{\nu_\tau}$ is negative for $Y_{\nu_\tau}
= 0$, implying that the Mikheyev-Smirnov-Wolfenstein resonance
occurs in the antineutrino
sector~\cite{Wolfenstein:1977ue,Mikheyev:1986gs}.

Given the effective Hamiltonian in Eq.~(\ref{eq:Hamiltonian}), one
immediately obtains the effective mixing angle
\begin{equation}\label{eq:angle}
\sin^2 2\theta_{\nu,\bar\nu} = \frac{\sin^2 2\theta}{\sin^2 2\theta + (\cos
  2\theta \pm E/E_{\rm r})^2} \;,
\end{equation}
where the upper sign refers to $\nu$ and the lower to $\bar\nu$. The
resonant energy $E_{\rm r} \equiv \Delta m^2/2|V_{\nu_\tau}|$ is
\begin{equation}\label{eq:resonant-energy}
E_{\rm r} = 3.25~{\rm MeV} \left(\frac{m_s}{10~{\rm
    keV}}\right)^2 \rho^{-1}_{14} \left|Y_0 - Y_{\nu_\tau}\right|^{-1} \; ,
\end{equation}
where $\rho_{14}$ is the matter density $\rho$ in units of
$10^{14}~{\rm g}~{\rm cm}^{-3}$ and $Y_0 \equiv (1 - Y_{e} -
2Y_{\nu_e})/4$. As indicated by Eq.~(\ref{eq:angle}), the mixing
angle $\theta_\nu$ for the whole energy range is always suppressed
by matter effects, while $\theta_{\bar\nu}$ can be resonantly
enhanced for $E \sim E_{\rm r} \cos 2\theta$. Note that the ``vacuum
limit" with $\theta_\nu \approx \theta_{\bar\nu} \approx \theta$ is
reached for large sterile neutrino masses $m_s \gg 10~{\rm keV}$,
while the ``medium limit'' with reduced mixing angle $\theta_\nu
\approx \theta_{\bar\nu} \approx (E_{\rm r}/E) \theta$ is obtained
for small masses $m_s \ll 10~{\rm keV}$. For intermediate masses, we
have a resonance in the antineutrino sector.

Sterile neutrinos are produced in the SN core via oscillations and
collisions of tau neutrinos. If the effective mixing angles
$\theta_\nu$ and $\theta_{\bar\nu}$ are small enough, $\nu_s$ and
$\bar\nu_s$ can escape from the core immediately after production.
Since the mixing angle of antineutrinos is always larger than that
of neutrinos, the emission rate of antineutrinos exceeds that of
neutrinos. Consequently, a $\nu_\tau$-$\bar\nu_\tau$ asymmetry
arises from these different emission rates. As we can observe from
Eq.~(\ref{eq:resonant-energy}), it might turn out that a relatively
large $\nu_\tau$-$\bar\nu_\tau$ asymmetry is achieved, i.e.,
$Y_{\nu_\tau} \to Y_0$, so as to drive the resonant energy $E_{\rm
r}$ to infinity, leading to the ``vacuum limit'' even for small
masses. Put another way, the $\nu_\tau$-$\bar\nu_\tau$ asymmetry
seems to develop in the direction of reducing the matter effects,
resulting in restrictive SN limits on the vacuum mixing angle
$\theta$ for both large and small sterile neutrino masses. It has
been argued~\cite{Abazajian:2001nj} that there exists a stationary
state with $Y_{\nu_\tau} = Y_0$, which can be achieved rapidly and
thus validates the ``vacuum limit" and restrictive bounds on
$\theta$.

However, the state with $Y_{\nu_\tau} = Y_0$ cannot be stationary,
because the population of $\nu_\tau$ at this moment is larger than
that of $\bar\nu_\tau$, implying that more neutrinos than
antineutrinos are ready to be emitted and thus the condition
$Y_{\nu_\tau} = Y_0$ breaks down. Furthermore, it is even possible
that the emission rate of $\nu_s$ exceeds that of $\bar\nu_s$, since
the effective mixing angle $\theta_{\bar\nu}$ can be so large that
most of $\bar\nu_s$ are trapped in the core. Therefore one may
obtain a negative asymmetry $Y_{\nu_\tau} < 0$, driving the system
towards the ``medium limit.'' In the following sections, we shall
examine how the $\nu_\tau$-$\bar\nu_\tau$ asymmetry actually
develops, and explore its implications on the anomalous energy-loss
rate of the SN core and thus the SN bounds on sterile neutrinos.

%%%%%%%%%%%%%%%%%%%%%%%%%%%%%%%%%%%%%%%%%%%%%%%%%%%%%%%%%%%%%%%%%%%%%%%
\section{Stationary state}                            \label{sec:state}
%%%%%%%%%%%%%%%%%%%%%%%%%%%%%%%%%%%%%%%%%%%%%%%%%%%%%%%%%%%%%%%%%%%%%%%

\subsection{Weak-damping limit}

The matter density of SN cores is so high that both neutrino
oscillations and frequent collisions with background particles are
important. An elegant method to treat neutrino flavor conversions in
this case is to implement the matrix of occupation numbers
$(\rho_{\bf p})_{ij} \equiv \langle b^\dagger_j({\bf p}) b_i({\bf
p})\rangle$, where $b_i({\bf p})$ denotes the annihilation operator
for a neutrino of flavor $i$ and momentum ${\bf p}$, and to derive
the non-Abelian Boltzmann equations of $\rho_{\bf p}$
\cite{Stodolsky:1986dx,Raffelt:1992uj,Sigl:1992fn}. The diagonal
elements $(\rho_{\bf p})_{ii}$ are the usual occupation numbers
$f^i_{\bf p}$, while the off-diagonal ones encode the phase
information. The analogous definition for antineutrinos is
$(\bar\rho_{\bf p})_{ij} \equiv \langle d^\dagger_i({\bf p})
d_j({\bf p})\rangle$ with $d_i({\bf p})$ being the annihilation
operator for an antineutrino of flavor $i$ and momentum ${\bf p}$.
In general, the equations of motion for $\rho_{\bf p}$ and
$\bar\rho_{\bf p}$ are complicated by the nonlinear nature of the
collision integrals.

For keV-mass sterile neutrinos, the problem can be much simplified by
taking the weak-damping limit, which is usually valid in SN cores.
To be more explicit, we estimate the neutrino oscillation length in
matter
\begin{equation}
\lambda_{\rm osc} \alt 0.7~{\rm cm} \left(\frac{E}{30~{\rm
    MeV}}\right) \left(\frac{10^{-4}}{\sin 2\theta}\right)
\left(\frac{10~{\rm keV}}{m_s}\right)^2
\end{equation}
and the mean free path of tau neutrinos
\begin{equation}
\lambda_{\rm mfp} = \frac{1}{N_{\rm B} \sigma_{\nu N}} \approx
1.1\times 10^3~{\rm cm} \left(\frac{30~{\rm MeV}}{E}\right)^2
\rho^{-1}_{14} \;,
\end{equation}
where $\sigma_{\nu N} \sim G^2_{\rm F} E^2/\pi$ is the cross section
of neutrino-nucleon scattering via the neutral-current interaction.
The weak-damping limit is $\lambda_{\rm osc} \ll \lambda_{\rm mfp}$,
meaning that the active-sterile neutrino oscillations take place
many times before a subsequent collision of active neutrinos with
the nucleons. It deserves mention that the weak-damping limit is
violated for smaller masses $m_s \ll 1~{\rm keV}$, however, the
effective neutrino mixing angle in this case is highly suppressed by
matter effects and thus the energy-loss rate is negligibly small. We
shall always assume the weak-damping limit in the mass range of our
interest.

\subsection{Emission rates}

In the weak-damping limit, $\rho_{\bf p}$ is averaged over many
cycles of oscillations and then can be parametrized by the neutrino
occupation numbers $f^\tau_E$ and $f^s_E$ with the neutrino energy
$E = |{\bf p}|$ and similar for antineutrinos~\cite{Raffelt:1992bs}.
Note that we consider a homogenous and isotropic ensemble of active
neutrinos and antineutrinos, which are trapped in the SN core and
stay in thermal equilibrium with ambient matter, so the momentum
direction is irrelevant. Assuming that sterile neutrinos freely
escape, we obtain the evolution equation of the $\nu_\tau$ density
\begin{equation}\label{eq:number-rate}
\dot{N}_{\nu_\tau} = -\frac{1}{4} \sum_a \int \frac{E^2 {\rm
    d}E}{2\pi^2} s^2_{2\theta_\nu} \int \frac{{E^\prime}^2 {\rm
    d}E^\prime}{2\pi^2} W^a_{E^\prime E} f^\tau_{E^\prime} \; ,~~
\end{equation}
where $a$ denotes the target particle, and $W^a_{E^\prime E}$ the
transition probability for $\nu_\tau(E^\prime) + a \to \nu_\tau(E) +
a$ via the neutral-current interaction. In a similar way, we can
derive the evolution equation of the $\bar\nu_\tau$ number density,
involving the mixing angle $\theta_{\bar\nu}$, the occupation number
$f^{\bar\tau}_E$ and the transition probability $\bar{W}^a_{E^\prime
E}$. However, only the neutrino energies ensuring $\theta_\nu,
\theta_{\bar\nu} \alt \theta_c$ are taken under the integration,
where the critical mixing angle $\theta_c \sim 10^{-2}$ can be
estimated by requiring the mean free path of sterile neutrinos to
equal the core radius $R \sim 10~{\rm km}$. On the other hand, the
transition probabilities $W^a_{E^\prime E}$ and ${\bar
W}^a_{E^\prime E}$ of neutrino-nucleon scattering dominate over
those of neutrino-electron scattering if the sterile neutrino mass
is lying in the range of $1~{\rm keV} \alt m_s \alt 100~{\rm keV}$.
Hence we consider only the neutral-current $\nu$-$N$ scattering in
the leading-order approximation.

\subsection{Degeneracy parameter}

In order to describe the $\nu_\tau$-$\bar\nu_\tau$ asymmetry, we
assume the occupation number $f^\tau_E(t) =
\left[\exp(E/T-\eta(t))+1\right]^{-1}$ for $\nu_\tau$ and
$f^{\bar\tau}_E(t) = \left[\exp(E/T+\eta(t))+1\right]^{-1}$ for
$\bar\nu_\tau$, where $T$ is the temperature of the SN core and
$\eta(t)$ is the degeneracy parameter. In the absence of
active-sterile neutrino mixing, the spectra of $\nu_\tau$ and
$\bar\nu_\tau$ just follow the Fermi-Dirac distribution with a
vanishing chemical potential, namely $\eta = 0$ at $t = 0$.
Substituting the occupation numbers into Eq.~(\ref{eq:number-rate}),
taking account of the neutral-current $\nu_\tau$-$N$ scattering and
subtracting the corresponding equation for antineutrinos, one
arrives at
\begin{equation}\label{eq:evolution}
\frac{{\rm d}\eta(t)}{{\rm d}t} = \frac{N_{\rm B}G^2_{\rm F}
s^2_{2\theta} T^2}{4\pi} \left[ \mathscr{F}_{\bar\nu}(\eta) -
\mathscr{F}_\nu(\eta)\right] \mathscr{G}^{-1}(\eta)\; ,
\end{equation}
where the relevant functions are defined as follows
\begin{eqnarray}\label{eq:function}
\mathscr{F}_{\bar\nu}(\eta) &=& \int^\infty_0
\frac{x^4}{e^{x+\eta}+1} \frac{1-\mathscr{B}(x,x_{\rm
    r}\epsilon^-,x_{\rm
    r}\epsilon^+)}{s^2_{2\theta}+(c_{2\theta}-x/x_{\rm r})^2}{\rm d}x
\; , \nonumber  \\
\mathscr{F}_{\nu}(\eta) &=& \int^\infty_0 \frac{x^4}{e^{x-\eta}+1}
\frac{\Theta(x+x_{\rm r}\epsilon^-)}{s^2_{2\theta} + (c_{2\theta} +
  x/x_{\rm r})^2} {\rm d}x \;,~~~~~~
\end{eqnarray}
and $\mathscr{G}(\eta) \equiv {\rm d}[F_2(\eta) - F_2(-\eta)]/{\rm
d}\eta$ with $F_2(\eta)$ being the Fermi-Dirac integral of order
two. In addition, we have introduced $x_{\rm r} \equiv E_{\rm r}/T$,
$\epsilon^\pm \equiv s_{2(\theta_c \pm
  \theta)}/s_{2\theta_c}$, and the box function $\mathscr{B}(x,a,b)
\equiv \Theta(x-a) - \Theta(x-b)$, which equals one for $x\in [a,
b]$ and vanishes otherwise. Here $\Theta(x)$ denotes the unit step
function, i.e., $\Theta(x) = 0$ for $x < 0$ and $\Theta(x) = 1$ for
$x \geq 0$. Note that Eq.~(\ref{eq:function}) has been cast into a
compact form so as to include both $\theta \leq \theta_c$ and
$\theta > \theta_c$ cases.

Taking typical values of the matter density $\rho_{14} = 3.0$ and
the core temperature $T = 30~{\rm MeV}$, we can rewrite
Eq.~(\ref{eq:evolution}) as $\dot{\eta}(t) = \tau^{-1}_0
\mathscr{H}(\eta)$ with $\tau_0 = 1~{\rm s}
\left(10^{-8}/s^2_{2\theta}\right)$ and $\mathscr{H}(\eta) = \left[
  \mathscr{F}_{\bar\nu}(\eta) - \mathscr{F}_\nu(\eta)\right]
\mathscr{G}^{-1}(\eta)$. The time evolution of the degeneracy
parameter $\eta(t)$ depends crucially on the initial difference
between neutrino and antineutrino emission rates, i.e.,
$\tau^{-1}_0\mathscr{H}(0)$, as well as the evolution of
$\mathscr{H}(\eta)$ with respect to $\eta$. In Fig.~1, we show the
initial rate $\dot{\eta}(0) = \tau^{-1}_0\mathscr{H}(0)$ in the
$(\sin^2 2\theta, m_s)$ plane, where the blank region with ``$+$"
denotes a strong excess of $\bar\nu_s$ over $\nu_s$ emission while
that with ``$-$" represents the opposite case. The cyan regions on
the left-hand and right-hand side indicate $\dot{\eta}(0)>0$ and
$\dot{\eta}(0)<0$, respectively, but the magnitude of
$|\dot{\eta}(0)|$ is extremely small for both cases. The reason for
the former case is just that the effective mixing angle is too
small, and for the latter case is that most of $\nu_s$ and
$\bar\nu_s$ are trapped.

%%%%%%%%%%%%%%%%%%%%%%%%%%%%%% FIG 1 %%%%%%%%%%%%%%%%%%%%%%%%%%%%%%%%
\begin{figure}[t]
\includegraphics[scale=0.8]{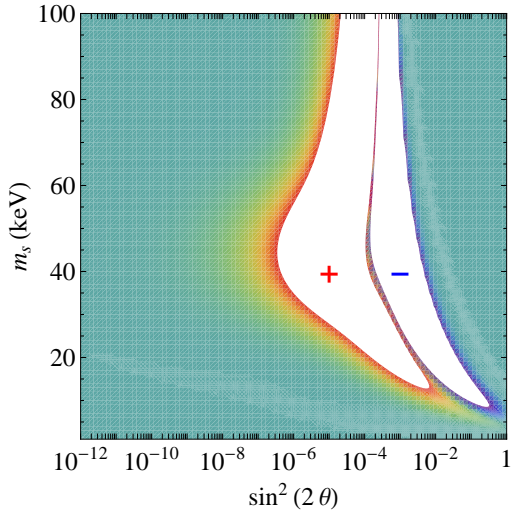}
\vspace{-0.1cm} \caption{Density plot of the initial rate
$\dot{\eta}(0) = \tau^{-1}_0 \mathscr{H}(0)$, where the blank region
with ``$+$" stands for a strong excess of $\bar{\nu}_s$ over
$\nu_s$ emission while that with ``$-$" is the other way round.
The narrow strip indicates a rapid and continuous transition between
these two regions.}
\end{figure}
%%%%%%%%%%%%%%%%%%%%%%%%%%%%%%%%%%%%%%%%%%%%%%%%%%%%%%%%%%%%%%%%%%%%%

If the initial emission rate of $\bar\nu_s$ is larger than that of
$\nu_s$, i.e., $\dot{\eta}(0) > 0$, the degeneracy parameter
increases from zero to a positive stable value $\eta^*$, at which
the emission rates of $\nu_s$ and $\bar\nu_s$ are equal, namely
$\mathscr{F}_{\bar\nu}(\eta^*) = \mathscr{F}_\nu(\eta^*)$ or
equivalently $\mathscr{H}(\eta^*) = 0$. In this case, we are finally
left with a positive $\nu_\tau$-$\bar\nu_\tau$ asymmetry when such a
stationary state is reached. If the initial emission rate of $\nu_s$
exceeds that of $\bar\nu_s$, i.e., $\dot{\eta}(0) < 0$, the
degeneracy parameter decreases from zero to a negative stable value
$\eta^*$. At this moment, we have equal neutrino and antineutrino
emission rates as well, but a negative $\nu_\tau$-$\bar\nu_\tau$
asymmetry.

The important point here is feedback of the established
$\nu_\tau$-$\bar\nu_\tau$ asymmetry or a finite degeneracy
parameter. In the case of $\dot{\eta}(0) > 0$, a positive $\eta$
suppresses the $\bar\nu_\tau$ population and shifts the resonant
energy to a larger value. The combined result is just to reduce the
$\bar\nu_s$ emission rate. Meanwhile, the population of $\nu_\tau$
is accordingly increased and the mixing angle $\theta_\nu$ becomes
less suppressed, enhancing the $\nu_s$ emission rate. Similar
arguments apply to the case of $\dot{\eta}(0) < 0$. It is also
possible that $\dot{\eta}(0) = 0$, which lies in the narrow strip in
Fig.~1, the system remains in its initial state with a vanishing
$\nu_\tau$-$\bar\nu_\tau$ asymmetry. Hence we have no feedback
effect in this special case.
%%%%%%%%%%%%%%%%%%%%%%%%%%%%%% FIG 2 %%%%%%%%%%%%%%%%%%%%%%%%%%%%%%%%
\begin{figure}[t]
\includegraphics[scale=0.8]{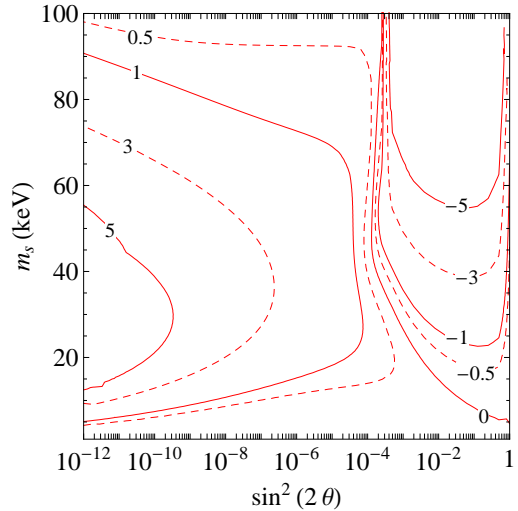}
\vspace{-0.1cm} \caption{Contour plot of the asymptotic
degeneracy parameter $\eta^*$, which is determined by
$\mathscr{F}_{\bar\nu}(\eta^*) = \mathscr{F}_\nu(\eta^*)$, implying
equal neutrino and antineutrino emission rates.}
\end{figure}
%%%%%%%%%%%%%%%%%%%%%%%%%%%%%%%%%%%%%%%%%%%%%%%%%%%%%%%%%%%%%%%%%%%%%
%%%%%%%%%%%%%%%%%%%%%%%%%%%%%% FIG 3 %%%%%%%%%%%%%%%%%%%%%%%%%%%%%%%%
\begin{figure}[t]
\includegraphics[scale=0.8]{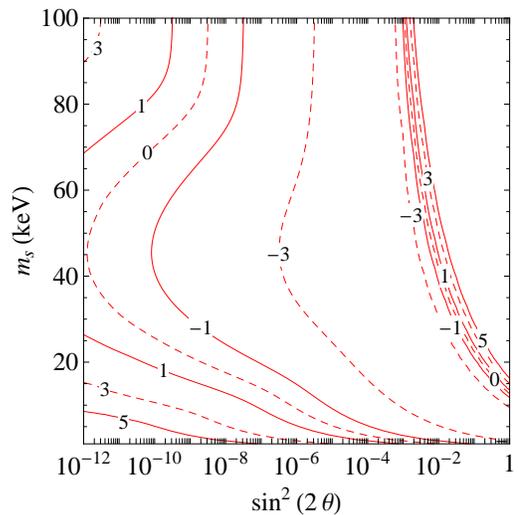}
\vspace{-0.1cm} \caption{Contour plot of the estimated timescale
$\tau = \eta^*/[\tau^{-1}_0 \mathscr{H}(0)]$ to reach the stationary
 state, where the contours are labeled by the logarithmic values
 $\log(\tau/1~{\rm s})$.}
\end{figure}
%%%%%%%%%%%%%%%%%%%%%%%%%%%%%%%%%%%%%%%%%%%%%%%%%%%%%%%%%%%%%%%%%%%%%

In Fig.~2, the degeneracy parameter $\eta^*$ of the stationary state
has been solved from $\mathscr{F}_{\bar\nu}(\eta) =
\mathscr{F}_\nu(\eta)$, no matter whether such a stationary state
can be reached. The large values of $\eta^*$ appear in the region
where the initial emission rate of~$\nu_s$ is significantly
different from that of~$\bar\nu_s$. This condition can be satisfied
for (1)~small mixing angles and intermediate masses, when the mixing
angle $\theta_{\bar\nu}$ is resonantly enhanced so that
$\mathscr{F}_{\bar\nu}(0) \gg \mathscr{F}_\nu(0)$; (2)~large mixing
angles and large masses, when more $\bar\nu_s$ than $\nu_s$ are
trapped in the core such that $\mathscr{F}_{\bar\nu}(0) \ll
\mathscr{F}_\nu(0)$. However, it is obvious that the $\nu_s$ and
$\bar\nu_s$ emission rates in both cases are extremely small.

The timescale for the system to achieve the stationary state can be
determined by numerically solving Eq.~(\ref{eq:evolution}). For a
rough estimate, we take $\dot{\eta}(t) \sim \dot{\eta}(0) =
\tau^{-1}_0 \mathscr{H}(0)$ and then obtain the timescale $\tau =
\eta^*/[\tau^{-1}_0 \mathscr{H}(0)]$. In Fig.~3, we show the
estimated timescale $\tau$ in the $(\sin^2 2\theta, m_s)$ plane. In
the mass range $1~{\rm keV} \alt m_s \alt 10~{\rm keV}$, where
sterile neutrinos can be warm dark matter, the timescale is larger
than the neutrino diffusion time $\tau_{\rm d} = 1~{\rm s}$ for
small mixing angles $\sin^2 2\theta \alt 10^{-6}$. For larger mixing
angles, the stationary state can be achieved within a fraction of a
second, but $\eta^*$ in this case is quite small as shown in Fig.~2.
Hence we expect that the feedback effects are negligible. In this
connection, the most interesting parameter space should be $20~{\rm
keV} \alt m_s \alt 80~{\rm keV}$ and $10^{-9} \alt \sin^2 2\theta
\alt 10^{-4}$, where both sizable $\eta^*$ and $\tau < \tau_{\rm d}$
are expected. Since $|{\mathscr H}(\eta)|$ decreases from the
initial value $|\mathscr{H}(0)|$ to zero, as the absolute value of
the degeneracy parameter $|\eta(t)|$ increases from zero to
$|\eta^*|$, the relaxation time may be underestimated. However, this
rough estimate has already shown the main features of the relaxation
timescale.

%%%%%%%%%%%%%%%%%%%%%%%%%%%%%%%%%%%%%%%%%%%%%%%%%%%%%%%%%%%%%%%%%%%%%
\section{Supernova bounds}                       \label{sec:snbounds}
%%%%%%%%%%%%%%%%%%%%%%%%%%%%%%%%%%%%%%%%%%%%%%%%%%%%%%%%%%%%%%%%%%%%%

%%%%%%%%%%%%%%%%%%%%%%%%%%%%%% FIG 4 %%%%%%%%%%%%%%%%%%%%%%%%%%%%%%%%
\begin{figure}[t]
\includegraphics[scale=0.8]{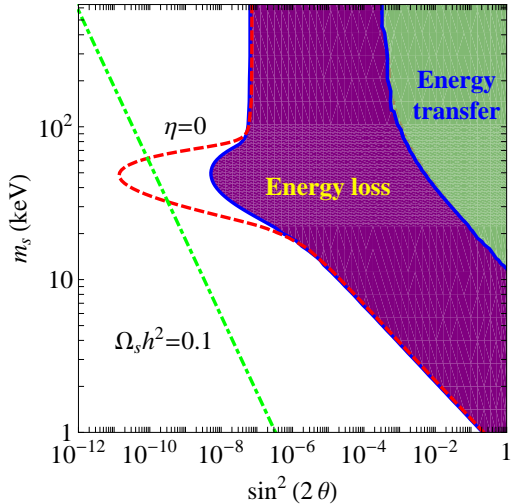}
\vspace{-0.1cm} \caption{Supernova bound on sterile neutrino masses
$m_s$ and mixing angles $\theta$, where the purple region is
excluded by the energy-loss argument while the green one by the
energy-transfer argument. The excluded region will be extended to
the dashed (red) line if the build-up of degeneracy parameter is
ignored, i.e., $\eta(t) = 0$. The dot-dashed (green) line represents
the sterile neutrinos as dark matter with the correct relic
abundance $\Omega_s h^2 = 0.1$. }
\end{figure}
%%%%%%%%%%%%%%%%%%%%%%%%%%%%%%%%%%%%%%%%%%%%%%%%%%%%%%%%%%%%%%%%%%%%%
The emission of sterile neutrinos $\nu_s$ and $\bar\nu_s$ may cause
rapid energy losses from the SN cores, which can significantly
shorten the duration of neutrino signals~\cite{Raffelt:1990yz}. In
order to avoid conflict with the observation of SN 1987A neutrinos,
we require the energy-loss rate per unit mass to be smaller than
$1.0\times 10^{19}~{\rm erg}~{\rm g}^{-1}~{\rm s}^{-1}$, which can
be translated into the volume energy-loss rate $\mathscr{E} <
3.0\times 10^{33}~{\rm erg}~{\rm cm}^{-3}~{\rm s}^{-1}$ for
$\rho_{14} = 3.0$. Given the emission rate of neutrinos in
Eq.~(\ref{eq:number-rate}) and the counterpart for antineutrinos,
the energy-loss rate is
\begin{equation}
\mathscr{E}(t) = \frac{N_{\rm B} G^2_{\rm F} s^2_{2\theta}
T^6}{8\pi^3} \left[\mathscr{R}_{\bar\nu}(\eta) +
\mathscr{R}_{\nu}(\eta)\right] \; ,
\end{equation}
with
\begin{eqnarray}\label{eq:energy}
\mathscr{R}_{\bar\nu}(\eta) &=& \int^\infty_0
\frac{x^5}{e^{x+\eta}+1} \frac{1-\mathscr{B}(x,x_{\rm
    r}\epsilon^-,x_{\rm
    r}\epsilon^+)}{s^2_{2\theta}+(c_{2\theta}-x/x_{\rm r})^2}{\rm d}x
\; , \nonumber  \\
\mathscr{R}_{\nu}(\eta) &=& \int^\infty_0 \frac{x^5}{e^{x-\eta}+1}
\frac{\Theta(x+x_{\rm r}\epsilon^-)}{s^2_{2\theta} + (c_{2\theta} +
  x/x_{\rm r})^2} {\rm d}x \;,~~~~~~
\end{eqnarray}
where the definitions of relevant parameters are given below
Eq.~(\ref{eq:function}). The energy-loss rate $\mathscr{E}(t)$ depends
on time through the degeneracy parameter $\eta(t)$, for which the time
evolution has been discussed in last section.

To constrain the sterile neutrino mass and mixing angle, we evaluate
the emission rate
\begin{equation}\label{eq:average}
\langle \mathscr{E}\rangle = \tau^{-1}_{\rm d} \int^{\tau_{\rm d}}_0
\mathscr{E}(t) {\rm d}t
\end{equation}
averaged over the neutrino diffusion timescale $\tau_{\rm d} =
1~{\rm s}$. Beyond the diffusion timescale, one may expect that all
active neutrinos have already diffused out of the core and thus the
emission of sterile neutrinos is physically meaningless. Our
strategy is to follow the time evolution of $\eta(t)$ for each point
in the $(\sin^2 2\theta, m_s)$ parameter space, and then calculate
the averaged energy-loss rate in Eq.~(\ref{eq:average}). We show in
Fig.~4 the contour plot of the averaged energy-loss rate $\langle
\mathscr{E}\rangle$ in the $(\sin^2 2\theta, m_s)$ plane, where the
purple region corresponds to $\langle \mathscr{E}\rangle > 3.0\times
10^{33}~{\rm erg}~{\rm cm}^{-3}~{\rm s}^{-1}$ and is thus excluded.

Some comments are in order. First, the initial energy-loss rate with
$\eta = 0$ is shown in Fig.~4 for comparison. Except for $20~{\rm
keV} \alt m_s \alt 100~{\rm keV}$ and small mixing angles, the
averaged energy-loss rate $\langle\mathscr{E}\rangle$ cannot be
distinguished from the initial one $\mathscr{E}(0)$. The reason is
that either the stationary state has not been reached within $1~{\rm
s}$, or the asymptotic value $\eta^*$ for the stationary state is
quite small. In the mass range $20~{\rm keV} < m_s < 100~{\rm keV}$,
the buildup of a $\nu_\tau$-$\bar\nu_\tau$ asymmetry is efficient
and strongly reduces the energy-loss rate.

Second, keV-mass sterile neutrinos can be produced in the early
universe and contribute as dark matter to the total energy density.
In a nonresonant production scheme without large primordial lepton
asymmetries, the relic sterile neutrino abundance can be estimated
as~\cite{Abazajian:2001nj,Asaka:2006nq}
\begin{equation}
\Omega_s h^2 \approx 0.3 \left(\frac{\sin^2
2\theta}{10^{-10}}\right) \left(\frac{m_s}{100~{\rm keV}}\right)^2
\; .
\end{equation}
The correct dark matter abundance $\Omega_s h^2 = 0.1$ is shown in
Fig.~4, where one can see that the masses around $m_s = 50~{\rm
keV}$ have been excluded by the SN bound if the degeneracy parameter
is assumed to be vanishing. In a realistic situation, this region is
retrieved because the energy-loss rate is reduced as the
$\nu_\tau$-$\bar\nu_\tau$ asymmetry builds up. However, the warm
dark matter range $1~{\rm keV} \alt m_s \alt 10~{\rm keV}$ is
essentially unconstrained. For sterile neutrinos of masses below
$1~{\rm keV}$, even maximal mixing is allowed, because the total
energy-loss rate is highly suppressed by matter effects.

Finally, our discussion was based on the energy-loss argument,
ignoring the sterile neutrinos trapped in the SN core. For this
reason, the top-right green region in Fig.~4, where both the sterile
neutrino mass and vacuum mixing angle are large, was not excluded.
However, the mean free path of these trapped sterile neutrinos is
much larger than that of $\nu_\tau$, so the energy transfer in the
SN core will be more efficient. The energy in a SN core is carried
by those trapped particles with the largest mean free path, a role
played here by the sterile states. Once more the neutrino burst
duration is shortened too much if the sterile mfp is larger than a
few times the one for $\nu_\tau$~\cite{Dolgov:2000jw}. Therefore,
the large mixing angle region is actually excluded in the spirit of
the energy-transfer argument.

%%%%%%%%%%%%%%%%%%%%%%%%%%%%%%%%%%%%%%%%%%%%%%%%%%%%%%%%%%%%%%%%%%%%%%
\section{Conclusions}
\label{sec:conclusions}
%%%%%%%%%%%%%%%%%%%%%%%%%%%%%%%%%%%%%%%%%%%%%%%%%%%%%%%%%%%%%%%%%%%%%%

Since keV-mass sterile neutrinos are a promising candidate for warm
dark matter, we have revisited the supernova bound on the sterile
neutrino masses and mixing angles by studying the case of
$\nu_\tau$-$\nu_s$ mixing in the SN core and requiring no excessive
energy losses induced by sterile neutrinos. It turns out that the
warm dark matter range is essentially unconstrained, while sterile
neutrinos of masses around $50~{\rm keV}$ receive the most stringent
constraint, i.e., $\sin^2 2\theta \alt 4.0\times 10^{-9}$. For even
larger masses $m_s \agt 100~{\rm keV}$, the SN limit on the mixing
angles is $\sin^2 2\theta \alt 5.0\times 10^{-8}$, which is about
one order of magnitude weaker than that for $m_s \sim 50~{\rm keV}$.
It is the matter effects that render these constraints quite
different.

We have identified a mass range $20~{\rm keV} \alt m_s \alt 100~{\rm
keV}$ where a sizable $\nu_\tau$-$\bar\nu_\tau$ asymmetry can be
established due to the strong excess of $\bar\nu_s$ over $\nu_s$
emission or vice versa. The build-up of this asymmetry feeds back on
the emission rates, leading to a stationary state where the neutrino
and antineutrino emissions become equal. For proper mixing angles,
such a stationary state can be achieved within the neutrino
diffusion timescale $\tau_{\rm d} = 1~{\rm s}$. As a consequence,
the energy-loss rate will be significantly reduced, and thus the
bounds are relaxed.

As for the $\nu_\mu$-$\nu_s$-mixing case, our discussions about the
feedback effects are essentially applicable. However, the
charged-current interactions of $\nu_\mu$ and $\bar\nu_\mu$ should
be taken into account, and the change of $\nu_\mu$-$\bar\nu_\mu$
asymmetry will be redistributed between muon neutrinos and charged
muons. The $\nu_e$-$\nu_s$ mixing in SN cores is more involved
because of the large trapped electron number and high $\nu_e$
degeneracy. Besides energy loss, deleptonization by sterile neutrino
emission is an effect to be taken into account. This case requires a
dedicated investigation.

We have performed a ``single zone'' analysis by assuming a
homogenous and isotropic SN core with constant matter density and
temperature. This treatment should capture the dominant feedback
effect. However, the local variation of these quantities may modify
the final results, for example smearing out the resonance in the
energy-loss rate. Such a refinement is also left for future works.

%%%%%%%%%%%%%%%%%%%%%%%%%%%%%%%%%%%%%%%%%%%%%%%%%%%%%%%%%%%%%%%%%%%%%%
\section*{Acknowledgements} %%%%%%%%%%%%%%%%%%%%%%%%%%%%%%%%%%%%%%%%%%
%%%%%%%%%%%%%%%%%%%%%%%%%%%%%%%%%%%%%%%%%%%%%%%%%%%%%%%%%%%%%%%%%%%%%%
This work was partly supported by the Deutsche
Forschungsgemeinschaft under Grants No. TR-27 and No. EXC-153 and by
the Alexander von Humboldt Foundation.

%%%%%%%%%%%%%%%%%%%%%%%%%%%%%%%%%%%%%%%%%%%%%%%%%%%%%%%%%%%%%%%%%%%%%%
%%%  Bibliography  %%%%%%%%%%%%%%%%%%%%%%%%%%%%%%%%%%%%%%%%%%%%%%%%%%%
%%%%%%%%%%%%%%%%%%%%%%%%%%%%%%%%%%%%%%%%%%%%%%%%%%%%%%%%%%%%%%%%%%%%%%

%%%%%%%%%%%%%%%%%%%%%%%%%%%%%%%%%%%%%%%%%%%%%%%%%%%%%%%%%%%%%%%%%%%%%%
\end{document}